\documentclass[12pt]{iopart}

\usepackage{subfigure}
\usepackage{graphicx}
\usepackage{lineno}
\usepackage{iopams}  
\begin{document}

\section*{Statement of Provenance}
This is an author-created, un-copyedited version of an article accepted for publication in Measurement Science and Technology. IOP Publishing Ltd is not responsible for any errors or omissions in this version of the manuscript or any version derived from it. The definitive publisher-authenticated version is available online at http://dx.doi.org/10.1088/0957-0233/24/3/037001 .
\clearpage
\linenumbers
\title[Simple feed-through for coupling optical fibres into hp/hT systems]{Simple feed-through for coupling optical fibres into high pressure and temperature systems}

\author{Thomas Reinsch, Christian Cunow, J\"{o}rg Schr\"{o}tter and Ronny Giese}

\address{Helmholtz Centre Potsdam GFZ German Research Centre for Geosciences, Telegrafenberg, Potsdam, Germany}
\ead{Thomas.Reinsch@gfz-potsdam.de}
\begin{abstract}
A best practice guide for assembling and testing a simple and inexpensive system feeding an optical fibre into a high pressure and temperature environment is presented. A standard Swagelok type connector is tested together with different ferrule materials and a PEEK capillary tube as feed-through. The system proofed to seal an optical fibre during several pressure and temperature cycling experiments up to 500~bar and 180$^{\circ}$C.

\end{abstract}

\vspace{2pc}
\submitto{\MST}
\vspace{2pc}
\noindent{\it Keywords}: feed-through, fibre optic, seals

\maketitle

\section{Introduction}
In order to use fibre optic sensors in high pressure and temperature applications, a feed-through for the optical fibre must be used to connect the sensors with the readout unit. Several similar approaches are reported for low pressure and high pressure applications. Sealings are established by compression of a plastic material onto the fibre \cite{Ip1990} as well as thermal contraction of a shrinking tube \cite{Butterworth1998}. Some authors report soldering an optical fibre to a feed-through. Furthermore, epoxy resin is used to seal optical fibres \cite{Weiss1985,Bock1988,Bohdan2004,Cowpe2008}. Combinations of different methods were tested as well \cite{Nishi1984}. Unlike solder and resin sealings, compression sealings allow for removing or adjusting the length of the optical fibre to be sealed.   

It is known, that a vacuum seal can be established using a standard Swagelok type fitting in combination with a Teflon \cite{Abraham1998} or an aluminium \cite{Miller2001} ferrule. Within this study, this approach is modified an tested for the application at high pressures. Here, a PEEK capillary is used as feed-through for the optical fibre. The capillary is sealed using a standard Swagelok type connector. Conducting simple tensile tests, different ferrule materials are evaluated regarding the required torque to seal the fibre, the repeatability of the sealing and the re-usability of the ferrule and capillary components. Furthermore, optical attenuation measurements at different wavelengths are performed to monitor the influence of the sealing on the optical properties of the fibre. After selecting an appropriate ferrule material, the sealing is tested in a high pressure and temperature (hp/hT) vessel at cyclic pressure and temperature conditions.

\section{Experiments}
\begin{figure}

\centering
\includegraphics[width=0.8\textwidth]{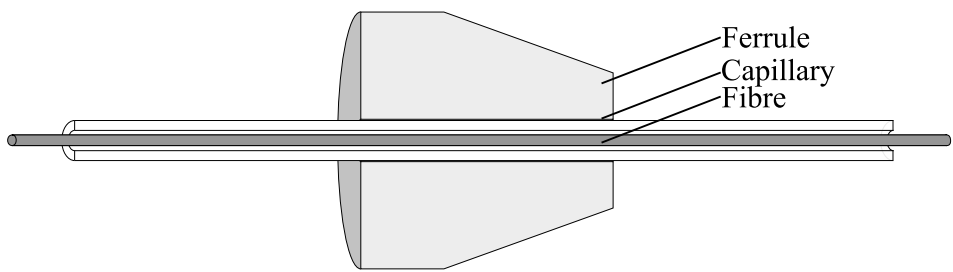}

\caption{Ferrule with capillary tube and optical fibre.}
\label{fig:ferrule}%
\end{figure}

A standard polyimide coated single mode optical fibre (8.3/125/155~$\mu$m) is used for the experiments. In order to seal the fibre, it is fed through a PEEK capillary tube (inner diameter 180~$\mu$m) with an additional ferrule (Figure \ref{fig:ferrule}). The ferrule is then inserted in to a standard 1/16~in Swagelok type fitting and the fitting is tightened using a torque handle. The sealing, therefore, is established due to the compression of the ferrule and hence of the PEEK capillary on the fibre.

\subsection{Tensile Test}
Before testing the sealing in a high pressure and high temperature vessel, an appropriate ferrule material is selected based on the results of a simple tensile test. Steel, PEEK, Teflon and Polyimide/Graphite (Pi/Gr 60$\%$/40$\%$) ferrules are tested. Therefore, the fitting is fixed in a bench vice and the fibre is pulled with a defined force, similar to the force acting on the sealing during a pressure experiment.

The force acting on the fibre during application of high pressure can be calculated according to the relation:
\begin{equation}
F=pA
\label{eq:force}
\end{equation}
where $p$ is the applied pressure and $A$ is the cross-sectional area of the fibre. Applying 500~bar to a 155~$\mu$m fibre, for example, a force of approximately 1~N acts on the fibre. Using a torque handle, different torque values are applied to the Swagelok fitting. Afterwards, the fibre is pulled with a force of 9~N, i.e.\ 900$\%$ safety. For every torque value, the optical attenuation is monitored using optical time domain reflectometry measurements at standard telecommunication wavelengths of 1310 and 1550~nm with a Wavetek MTS6000. Therefore, a fibre optic extension of about 4~km is spliced to the tested fibre.

\subsection{hp/hT Test}
From the results of the tensile test, a Pi/Gr ferrule is chosen for high pressure and temperature testing (see Section \ref{sec:res}). A Swagelok type fitting is mounted to a high pressure and temperature vessel \cite{Milsch2008} in order to test its performance under cyclic pressures and temperature conditions. Using a pressurizing oil, pressures up to $500~\mathrm{bar}$ and temperatures up to 180$^{\circ}$C are applied. Again, the optical attenuation is monitored at 1310 and 1550~nm. The schedule of the experiments is listed in Table \ref{tab:schedule}. 
\begin{table}
\caption{\label{tab:schedule}Schedule of the pressure and temperature cycling experiments. RT refers to room temperature conditions. The time is given approximately, as the chamber is not actively cooled and changing the temperature, therefore, takes several hours.}
\footnotesize\rm
\centering
\begin{indented}
\item[]\begin{tabular}{r|rr}
\br
Time&Applied Temperature&Applied Pressure\\
h&$^{\circ}$C&bar\\
\mr
0	&	RT	&	0	\\
19	&	RT	&	450	\\
76	&	RT	&	500	\\
24	&	RT	&	0	\\
24	&	RT	&	200	\\
24	&	150	&	400	\\
21	&	170	&	100	\\
24	&	50	&	0	\\
120	&	170	&	400	\\
48	&	180	&	500	\\
10	&	RT	&	0	\\

\br
\end{tabular}
\end{indented}

\end{table}

\section{Results}\label{sec:res}

\subsection{Tensile Test}
The torque, necessary to seal the fibre within the fitting, is listed in Table \ref{tab:material} for the different ferrule materials. It is also indicated if the ferrule can be reused after the experiment, i.e.\ if it was possible to remove it from the capillary tube. Different ferrule materials exhibit a different mechanical behaviour. Except for the steel ferrule, all ferrule materials passed the tensile test.

\begin{table}
\caption{\label{tab:material}List of tested ferrule materials. A force of 9~N is applied during the tensile test. The required torque to seal the fibre within the fitting is given together with information on the repeatability of the experiments and the re-usability of the sealing.}
\footnotesize\rm
\centering
\begin{indented}
\item[]\begin{tabular}{r|rrr}
\br
Material& Torque &Repeatable&Reusable\\
&(Nm)&&\\
\mr
Steel&-&no&no\\
PEEK&4&yes&no\\
Teflon&2&yes&yes\\
Pi/Gr (60/40)&2&yes&yes\\
\br
\end{tabular}
\end{indented}

\end{table}

\begin{figure}
\centering
\subfigure[1310~nm]{\includegraphics[width=0.45\textwidth]{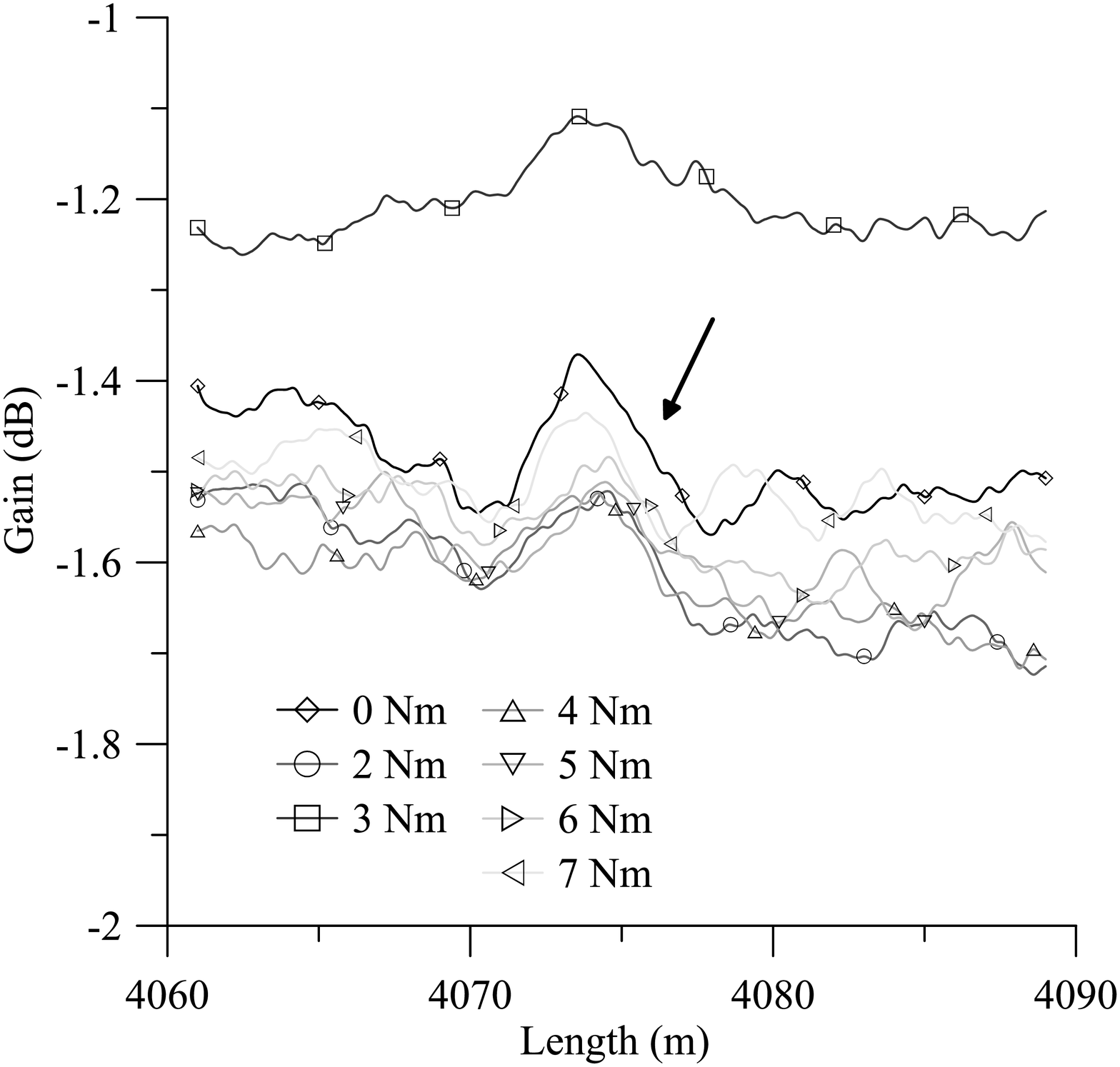}}\hspace{0.5cm}
\subfigure[1550~nm]{\includegraphics[width=0.45\textwidth]{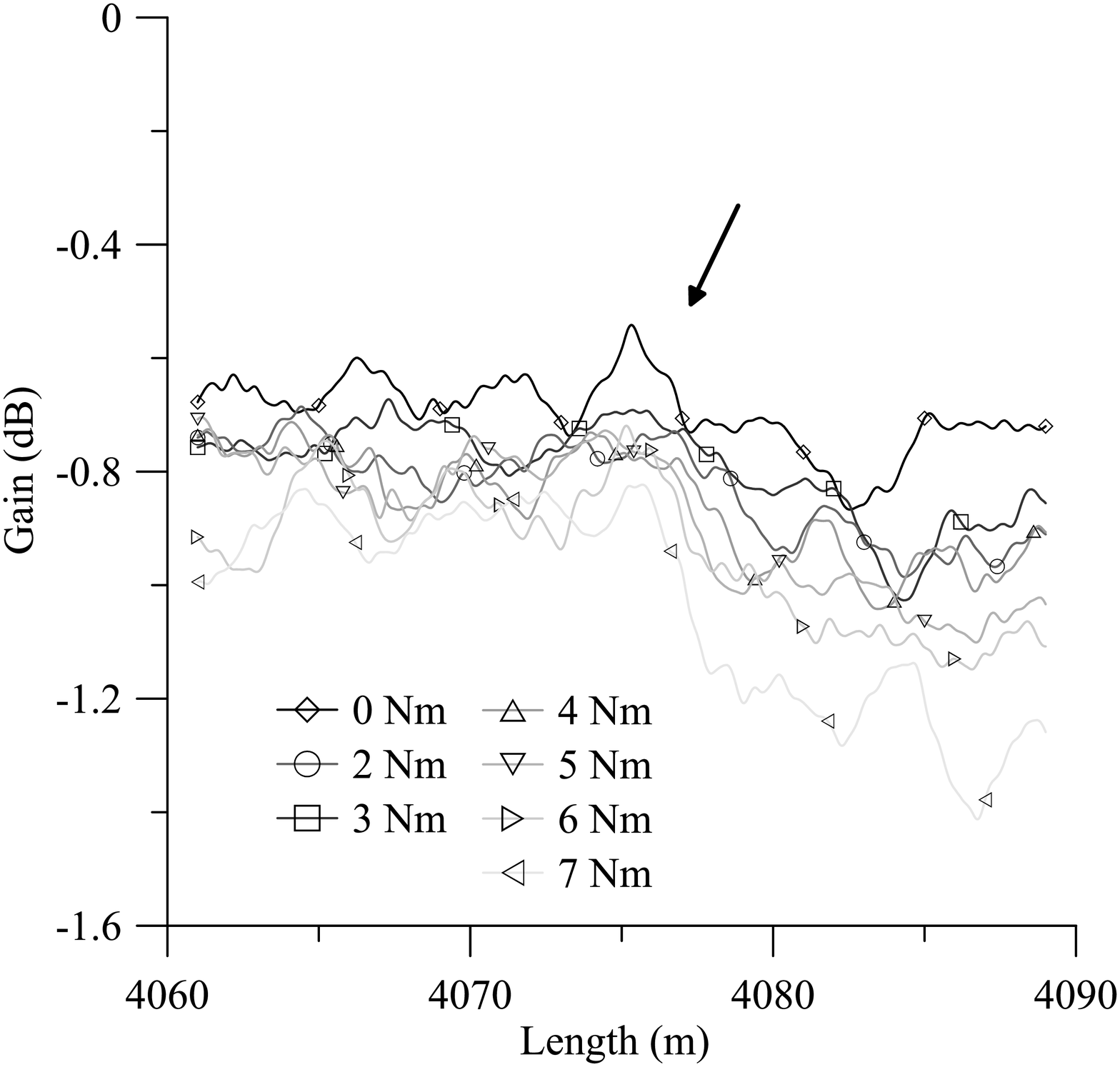}}
\caption{Gain at 1310 and 1550~nm for different torque values. The position of the sealing is indicated. The pulse length of the MTS6000 was 10~ns and the spatial resolution 4~cm. Here, a 2~m moving average is displayed.}%
\label{fig:wavetek}
\end{figure}

For different torque values, attenuation measurements were performed at 1310 and 1550~nm. Figure \ref{fig:wavetek} shows the results for a Pi/Gr ferrule and different torques. No influence can be observed at 1300~nm for different torque values. At 1550~nm, a slightly increasing attenuation is observed with increasing torque, indicating increasing mechanical stress on the optical fibre.

\subsection{hp/hT Test}
After performing the tensile tests, the fitting is mounted to the inside of a high pressure/high temperature vessel. It is fixed with a torque of 2.5~Nm. During the pressure and temperature cycling experiments, neither leaking of oil nor an influence on the measured attenuation can be observed.

\section{Discussion and Conclusions}

The presented sealing technique proved to be a simple, inexpensive and appropriate for a variety of applications in high pressure and temperature experimentation. Except for the steel ferrule, the different ferrule materials proved to be applicable for a high pressure sealing of an optical fibre during the tensile test. The presented tests show that the optical properties do not change significantly upon compression along the ferrule applying the required torque.

\ack
This work was performed within the framework GeoEn-Phase~2 project and funded by the Federal Ministry of Education and Research (BMBF, 03G0767A). The authors are grateful to S. Raab and E. Spangenberg for the fruitful discussions about different ferrule materials and their assistance during the high pressureand temperature experiments, as well as H. Milsch for his comments on the manuscript.

\section*{References}
\bibliographystyle{unsrt}
\bibliography{Feedthrough_MST}

\end{document}